\documentclass[11pt]{article}

\input epsf
\usepackage{latexsym}
\usepackage{amssymb}
\usepackage{amsmath}
\usepackage[dvips]{graphicx}

\begin{document}
\centerline{\Large \bf T-Duality and Penrose limits of spatially homogeneous }
\vspace{0.3cm}
\centerline{\Large \bf
and inhomogeneous cosmologies}
\vskip 2 cm

\centerline{Kerstin E. Kunze\footnote{E-mail: Kerstin.Kunze@physik.uni-freiburg.de} }

\vskip 0.3cm

\centerline{{\sl Physikalisches Institut,
Albert-Ludwigs-Universit\"at Freiburg, }}
\centerline{{\sl Hermann-Herder-Strasse 3,
D-79104 Freiburg, Germany. }}

\vskip 1.5cm

\centerline{\bf Abstract}

\vskip 0.5cm
\noindent
Penrose limits of inhomogeneous cosmologies admitting
two abelian Killing vectors and their abelian T-duals
are found in general. 
The wave profiles of the resulting plane waves 
are given for particular solutions.
Abelian and non-abelian T-duality are used as  solution generating techniques.
Furthermore, it is found that unlike in the case 
of abelian T-duality, non-abelian T-duality and taking the Penrose
limit are not commutative procedures.

\vskip 1cm

\section{Introduction}
The low energy limit of string-/M theory admits 
a variety of cosmological solutions.
In four dimensions, these string cosmologies differ from cosmologies 
derived from general relativity
due to the presence of scalar fields and  
form fields (see for example, \cite{str-co}).
In analogy to standard cosmology, string cosmologies, as well,
generically have an initial space-time singularity.
Close to any singularity the low energy approximation breaks down 
and the full string-/M theory is needed. 
However, in general it is not clear how to relate 
solutions of the low energy limit to 
exact string solutions and if this is at all possible.
Plane waves are known examples of exact classical string vacua  
\cite{pla-ex}. This means that they are exact to all orders  
in the string tension $\alpha'$. 
Recent developments in string-/M theory have let to  renewed 
interest in an argument by Penrose \cite{penrose}
showing that all space-times,
locally, in the neighbourhood of a null geodesic have a plane
wave as a limit \cite{M-pen, blau}. Therefore, in the Penrose
limit,
any space-time
can be related to an exact classical string vacuum. 
For some of the plane wave backgrounds 
descriptions in terms of (solvable)
conformal field theories have been found, 
which determine the spectrum of string 
excitations and their scattering amplitudes \cite{cft}. 
Recently, superstrings in plane wave backgrounds have been also
discussed \cite{sup-pw}.

Duality transformations relate different string backgrounds.
Therefore different string cosmologies can be connected by
symmetries. Abelian T-duality allows to transform
backgrounds admitting at least one abelian 
isometry into another background of this type.
The transformation changes the metric, antisymmetric
tensor field and the dilaton while keeping the
abelian isometry of the background \cite{bu, giv,berg}.
Similarly, non-abelian T-duality transforms
backgrounds with non-abelian isometries.
However, in this case the non-abelian isometry might be 
lost during the transformation. Therefore
backgrounds without any kind of symmetry might be related 
to ones admitting non-abelian isometries \cite{oq, grv, aal}.
With the extra fields being constants (or zero)
general relativity is a particular solution of low energy string theory.
Most of the solutions of general relativity admit some 
kind of abelian or non-abelian symmetries. 
Therefore using abelian and/or non-abelian T-duality
new solutions to string cosmology can be found. This has led
already to a multitude of solutions \cite{str-co}.
However, in addition to finding new solutions of 
string cosmology it should be noted that  these symmetries can also be 
used as solution generating techniques within standard 
general relativity. 

Using T-duality transformations a given background can be connected to
a variety of different string cosmologies. In the Penrose limit 
all of these reduce to a plane wave space-time.
Therefore, it might be worthwile to see if the
resulting plane waves are connected by a T-duality tranformation,
or in other words, whether  taking the Penrose limit and
dualizing are commutative.

In the following, the Penrose limiting procedure and abelian
and non-abelian T-duality are briefly reviewed.

According to \cite{penrose} 
any D dimensional metric in the neighbourhood of a
segment of a null geodesic containing no conjugate points 
can be written as \cite{blau}
\begin{eqnarray}
ds^2=dudv+\alpha dv^2+\sum_i\beta_idvdx^i+C_{ij}dx^idx^j
\label{e1}
\end{eqnarray}
where $\alpha$, $\beta_i$ and $C_{ij}$ are functions
of all coordinates and $i,j=1,2,..,D-2$. 
Following Penrose the coordinates are rescaled 
by a constant factor $\Omega>0$, 
\begin{eqnarray}
u=\tilde{u}\;\;\;\;\;v=\Omega^2\tilde{v}\;\;\;\;\;\;x^i=\Omega\tilde{x}^i.
\label{resc}
\end{eqnarray}
Taking the limit $\Omega\rightarrow 0$ of $d\tilde{s}^2/\Omega^2$
gives the behaviour of the metric in the neighbourhood of
a null geodesic. In this case $\tilde{u}$ is 
an affine parameter.
G\"uven \cite{guev} extended the Penrose limit to include other
fields, such as gauge and scalar fields. In summary, for a scalar field,
e.g. the dilaton, $\phi$, the antisymmetric
tensor field $B=B_{MN}dX^M\wedge dX^N$ and the metric the  
behaviour in the Penrose limit is given  by
\begin{eqnarray}
\hat{\phi}&=&\lim_{\Omega\rightarrow 0}\phi(\Omega)\nonumber\\
\hat{B}&=&\lim_{\Omega\rightarrow 0}\Omega^{-2}B(\Omega)\nonumber\\
d\hat{s}^2&=&\lim_{\Omega\rightarrow 0}\Omega^{-2}ds^2(\Omega)
\label{penrl}
\end{eqnarray}
where the argument $\Omega$ denotes the rescaling of variables (\ref{resc}).

Duality symmetries relate different string backgrounds. Abelian T-duality is 
a symmetry with respect to an abelian Killing direction.
T-dualities are derived from the 
two-dimensional $\sigma$-model action 
given by,
\begin{eqnarray}
S=\frac{1}{4\pi}\int d^2z
\left\{
\partial X^M\left[G_{MN}(X)+B_{MN}(X)\right]
\bar{\partial}X^N+\frac{1}{2}R^{(2)}\phi(X)
\right\},
\label{sig}
\end{eqnarray}
where $M,N=0,...,d$, $X^M\equiv (t,X^m) (m=1,...,d)$,
are the string coordinates, $R^{(2)}$ is the scalar
curvature of the 2-dimensional worldsheet and $G_{MN}, B_{MN}$
and $\phi$ are functions of $X$.
Choosing coordinates $\{x^{\mu}\}=\{x^0,x^a\}$ such that the
abelian isometry acts by translation of $x^0\equiv \theta$
and all background fields are independent of $\theta$.
The T-duality transformation is found by gauging the Abelian isometry and
then introducing Lagrangian multipliers in order to keep the gauge 
connection flat. These Lagrangian multipliers are promoted to coordinates
in the dual space-time. Dual and original quantities are related as follows
\cite{bu} \cite{giv},
\begin{eqnarray}
G_{00}'&=&\frac{1}{G_{00}}, \hspace*{1cm} G_{0a}'=\frac{B_{0a}}{G_{00}}, \hspace*{1cm}
G_{ab}'=G_{ab}-\frac{G_{a0}G_{0b}+B_{a0}B_{0b}}{G_{00}},\nonumber\\
B_{0a}'&=&\frac{G_{0a}}{G_{00}}, \hspace*{1cm}
B_{ab}'=B_{ab}-\frac{G_{a0}B_{0b}+B_{a0}G_{0b}}{G_{00}} .
\label{AD1}
\end{eqnarray} 
The dilaton is shifted to 
\begin{eqnarray}
\phi'=\phi-\log G_{00}.
\label{AD2}
\end{eqnarray}

In \cite{oq} a T-duality transformation  for backgrounds with non-abelian
isometries was proposed. However, in \cite{grv} an example,
namely a Bianchi V cosmology, was given
for which this transformation does not lead to another
consistent string background since the 
(low energy) $\beta$ function equations are not satisfied.
In \cite{aal} it was shown that in the case
that the group of isometries of the background is not
semi-simple, which is the case for Bianchi V, a mixed gauge and
gravitational anomaly is present.   
However,  in \cite{moh} it was found that not all non-semi-simple 
groups lead to an anomaly.
Non-abelian duality transformations have been generalized
to Poisson-Lie T-duality which allows to find dual
space-times even with respect to the non-semi-simple groups 
that were excluded for non-abelian T-duality \cite{pois}.
However, here the focus will be on the standard
non-abelian T-duality procedure \cite{oq}.

In general, it is not possible to write
explicitly the gauge fixed action. Thus the dual fields
cannot be presented in a closed form as it was possible in the
abelian case (cf. eqs.(\ref{AD1}) and (\ref{AD2})) \cite{oq}.

In the following spatially homogeneous and simple inhomogeneous cosmologies
will be investigated. The corresponding metrics admit three or
two Killing vectors, respectively. Whereas the former admit non-abelian isometries, the
latter are abelian. Therefore the structure is rich enough to apply 
abelian and non-abelian T-duality. The observable universe on large
scales is well described by a Friedmann-Robertson-Walker universe which
is a particular case of a spatially homogeneous universe. However, with
a view to the question of initial conditions more general
cosmologies deserve further study as well. 
The spatially homogeneous models were first classified by Bianchi into nine
different types (cf. \cite{exact}).
Bianchi models I to VII, locally rotationally symmetric (LRS)
VIII and LRS IX have two-dimensional abelian subgroups.
Therefore these can be described in the same fashion as
spatially inhomogeneous space-times admitting two abelian Killing vectors.

\section{Abelian T-duality of $G_2$ cosmologies and the radial Penrose limit}
$G_2$ space-times admit two abelian Killing vectors. Thus spatial homogeneity
is broken along one spatial direction. In general these metrics can be written as 
\cite{grif}
\begin{eqnarray}
ds^2=2e^{-M}dudv-\frac{2e^{-U}}{Z+\bar{Z}}\left(dx+iZdy\right)\left(dx-i\bar{Z}dy\right),
\label{g2}
\end{eqnarray}
where $M$ and $U$ are real and $Z$ is a complex function of the two null coordinates
$u$ and $v$. Therefore these
space-times (\ref{g2}) are conveniently described in terms of a null tetrad.

Introducing coordinates $t=u-v$, $r=u+v$, say,  makes the line element
(\ref{g2}) similar to that of a cylindrical space-time. In that case,
$r$ could be interpreted as the radius of the cylinder.
Geodesics in cylindrical space-times have been investigated 
in connection with nonsingular solutions in \cite{ns}.
Although due to the presence of two abelian Killing vectors there are
two constants of motion in the set of geodesic equations the general
solution is not straightforward to find and one has to specialize to
certain types of geodesics. For radial geodesics the constants of motion are zero 
and explicit solutions can be found in closed form. 
Furthermore, the change to adapted null coordinates is not obvious. 
Therefore, in the following, only
Penrose limits around radial null geodesics will be investigated.

The limiting procedure of Penrose \cite{penrose}
can be applied along a segment  of a null geodesic without 
conjugate points. This means that the expansion of a congruence
of neighbouring null geodesics has to be finite.
For geodesics with tangent vector parallel to 
$n^{\mu}=e^{M/2}\partial_u$ the expansion 
is given by $\mu+\bar{\mu}=e^{M/2}\left(e^{-U}\right)_u /
e^{-U}$ 
and equivalently for those with tangent vector parallel to
$l^{\mu}=e^{M/2}\partial_v$ the expansion 
is given by $\rho+\bar{\rho}=-e^{M/2}\left(e^{-U}\right)_v /
e^{-U}$ \cite{grif}. Here $\mu$ and $\rho$ are
Newman-Penrose spin coefficients.   
Therefore assuming that $e^{M/2}$ and $e^{U}$ are bounded,
the Penrose limit (\ref{penrl}) of the metric (\ref{g2})
leads to a plane wave space-time with all functions just depending
on one of the null coordinates, say $u$.
However, in general the null coordinate will not be 
an affine parameter. Therefore in the following it is
assumed that after taking the Penrose limit 
a new null coordinate $u=\int e^{-M(\tilde{u})}d\tilde{u}$
has been introduced.
For plane waves, traveling in $u$ direction, 
the only nonvanishing null tetrad component of the Weyl tensor
is given by \cite{grif} 
\begin{eqnarray}
\Psi_4=\frac{Z_{uu}-U_uZ_u}
{Z+\bar{Z}}-2\frac{\left(Z_u\right)^2}{\left(Z+\bar{Z}\right)^2}.
\end{eqnarray}
The only nonvanishing tetrad component of the Ricci tensor is given by
\begin{eqnarray}
\Phi_{22}=\frac{1}{4}
\left[
2U_{uu}-\left(U_u\right)^2-4\frac{Z_u\bar{Z}_u}{\left(Z+\bar{Z}\right)^2}
\right].
\end{eqnarray}

In analogy with electromagnetism, $\Psi_4$ can be written as $\Psi_4=Ae^{i\alpha}$
where $A$ is the amplitude and
$\alpha$ the polarization of the gravitational wave \cite{grif}. Therefore 
$\Psi_4$ determines the profile of the wave. It is interesting to note that 
the Brinkmann form of the metric can be read off from
$\Psi_4$ and $\Phi_{22}$.
The Brinkmann form is given by
\begin{eqnarray}
ds^2=2dudV+\left(h_{11}X^2+2h_{12}XY+h_{22}Y^2\right)du^2-dX^2-dY^2,
\label{bri}
\end{eqnarray}
where $h_{ij}$ are functions of $u$ only. The Weyl and Ricci tensor components are 
given by \cite{grif}
\begin{eqnarray}
\Psi_4=\frac{1}{2}\left(h_{11}-h_{22}+2ih_{12}\right)
\;\;\;\;\;\;\;\;\;\;\;\;\;\;\;\;\;\;\;\;\;\;\;\;
\Phi_{22}=\frac{1}{2}\left(h_{11}+h_{22}\right).
\end{eqnarray}
Therefore calculating these quantities for the Einstein-Rosen form (\ref{g2})
allows to read off the profile of the gravitational wave, $h_{ij}$, in the 
Brinkmann form.

Assuming that the metric (\ref{g2}) describes a vacuum space-time
the following Brinkmann form for the resulting plane wave in the 
Penrose limit is obtained,
\begin{eqnarray}
h_{11}&=&-\frac{\left[\left(\frac{2e^{-U}}{Z+\bar{Z}}\right)^{\frac{1}{2}}\right]_{uu}}
{\left(\frac{2e^{-U}}{Z+\bar{Z}}\right)^{\frac{1}{2}}}
-\frac{1}{4}\frac{\left[\left(Z-\bar{Z}\right)_u\right]^2}
{\left(Z+\bar{Z}\right)^2}
\end{eqnarray}
\begin{eqnarray}
h_{12}&=&-\frac{i}{2}\frac{\left(Z-\bar{Z}\right)_{uu}-
U_u\left(Z-\bar{Z}\right)_u}{Z+\bar{Z}}
+i\frac{\left(Z_u\right)^2-\left(\bar{Z}_u\right)^2}
{\left(Z+\bar{Z}\right)^2}
\\
h_{22}&=&-\frac{\left[\left(\frac{Z+\bar{Z}}{2}e^{-U}\right)^{\frac{1}{2}}\right]_{uu}}
{\left(\frac{Z+\bar{Z}}{2}e^{-U}\right)^{\frac{1}{2}}}
+\frac{3}{4}\frac{\left[\left(Z-\bar{Z}\right)_u\right]^2}
{\left(Z+\bar{Z}\right)^2}.
\end{eqnarray}
Using that $\Phi_{22}=0$ in vacuum it follows that $h_{11}=-h_{22}$.

The abelian T-duality transformations (\ref{AD1})
take a particularly simple form in terms of the functions $U$ and $Z$ when
applied for $\phi=0$ and $B_{\mu\nu}=0$.
The function $M$ remains invariant under this duality transformation.
T-duality with respect to the Killing vector $\partial_x$ results in
\begin{eqnarray}
e^{-U'}&=&\frac{Z+\bar{Z}}{2},  \hspace*{1.7cm}
Z'=e^{-U}\nonumber\\
B_{xy}'&=&\frac{i}{2}\left(Z-\bar{Z}\right), \hspace*{1cm}
\phi'=-\ln\frac{2e^{-U}}{Z+\bar{Z}}.
\label{Bu_x}
\end{eqnarray}
The metric is diagonal and hence $h_{11}=\Psi_4+\Phi_{22}$ and 
$h_{22}=\Phi_{22}-\Psi_4$, which yields to
\begin{eqnarray}
h_{11}=-\frac{\left[\left(\frac{Z+\bar{Z}}{2}e^{U}\right)^{\frac{1}{2}}\right]_{uu}}
{\left(\frac{Z+\bar{Z}}{2}e^{U}\right)^{\frac{1}{2}}},
\hspace*{1cm}
h_{22}=-\frac{\left[\left(\frac{Z+\bar{Z}}{2}e^{-U}\right)^{\frac{1}{2}}\right]_{uu}}
{\left(\frac{Z+\bar{Z}}{2}e^{-U}\right)^{\frac{1}{2}}}.
\end{eqnarray}
Therefore if the seed metric (\ref{g2}) is diagonal then
the wave profile in the direction orthogonal to the 
Killing direction along which the T-duality transformation is taken 
remains invariant.

T-duality with respect to the Killing vector $\partial_y$ results in
\begin{eqnarray}
e^{-U'}&=&\frac{Z+\bar{Z}}{2Z\bar{Z}}, \hspace*{1.3cm}
Z'=e^{U}\nonumber\\
B_{xy}'&=&\frac{i}{2}\frac{Z-\bar{Z}}{Z\bar{Z}}, \hspace*{1cm}
\phi'=-\ln\left(\frac{2e^{-U}}{Z+\bar{Z}}Z\bar{Z}\right).
\label{Bu_y}
\end{eqnarray}
The wave profile is given by
\begin{eqnarray}
h_{11}=-\frac{\left[\left(\frac{Z+\bar{Z}}{2Z\bar{Z}}e^{-U}\right)^{\frac{1}{2}}\right]_{uu}}
{\left(\frac{Z+\bar{Z}}{2Z\bar{Z}}e^{-U}\right)^{\frac{1}{2}}}\;\;\;\;\;\;\;\;\;\;\;\;\;\;
\;\;\;\;\;\;\;\;
h_{22}=-\frac{\left[\left(\frac{Z+\bar{Z}}{2Z\bar{Z}}e^{U}\right)^{\frac{1}{2}}\right]_{uu}}
{\left(\frac{Z+\bar{Z}}{2Z\bar{Z}}e^{U}\right)^{\frac{1}{2}}}.
\end{eqnarray}
Again it is found that in the case of a diagonal seed 
metric the wave profile stays invariant
in the direction 
orthogonal to the Killing direction along which the 
T-duality transformation is taken.

It is interesting to note that different spatially homogeneous backgrounds can be
related to each other using
the abelian T-duality transformations (\ref{Bu_x})
and (\ref{Bu_y}). 
Isometries of spatially homogeneous metrics in four dimensions
are described by three space-like Killing vectors that form an
algebra. In total there are nine different types originally 
classified by Bianchi (cf. e.g.  \cite{exact}). 
They fall into two classes, A and B, 
according to whether the trace of the group structure constants 
vanishes or not. 
Bianchi types I, II, VI$_{-1}$,
VII$_0$, VIII and IX are of class A whereas Bianchi types
III, IV, V, VI$_{h}$ and VII$_h$
are of class B. 

Bianchi class A models can always be described
by a diagonal metric in the invariant basis, i.e.
$ds^2=dt^2-g_{ij}(t)\omega^i\omega^j$ where 
$\omega^i$ are the invariant basis one forms, satisfying
$d\omega^i=\frac{1}{2}C^i_{\;\;jk}\omega^j\wedge\omega^k$
and $C^i_{\;\;jk}$ are the group structure constants.
Furthermore the metric is assumed to be of the form
$g_{ij}={\rm diag}\left(a_1^2(t), a_2^2(t), a_3^2(t)\right)$.
A Bianchi type V background can also be described by
a diagonal metric. However, for Bianchi type IV a nondiagonal
metric is required. In order to investigate its behaviour under the 
duality transformations (\ref{Bu_x}) and (\ref{Bu_y}) the ansatz of 
Harvey and Tsoubelis \cite{ht} was used. Namely,
$\sigma^1=a_1\omega^1, \sigma^2=a_2\omega^2, 
\sigma^3=a_3f\omega^2+a_3\omega^3$,  where 
$a_i$ and $f$ are functions of the timelike variable, $t$, only.
$\sigma^i$ are the basis one forms in the orthormal frame,
$ds^2=\eta_{\mu\nu}\sigma^{\mu}\sigma^{\nu}$, with 
$\eta_{\mu\nu}$ the Minkowski metric. Harvey and Tsoubelis
found a solution, which, incidently, describes a plane wave, for 
$a_2=a_3$ \cite{ht}.
No spatially homogeneous background was found when the
T-duality transformations  (\ref{Bu_x}) and (\ref{Bu_y}) were applied
to backgrounds of Bianchi type VI$_h$ /  VII$_h$, namely to the 
Lukash type metric \cite{luk}.
Bianchi models I to VII, LRS VIII and LRS IX have two-dimensional abelian subgroups.
Therefore they can be written in the form of metric (\ref{g2}).

In Table 1 those Bianchi models  
are given for which an abelian T-duality transformation leads 
to another spatially homogeneous background.
It was explicitly checked that the resulting dilaton and 
antisymmetric tensor field are consistent with the spatial
homogeneity of the background. It was found that for the
seed metrics of  Bianchi type IV, V and VI$_{-1}$ the resulting
dilaton acquires a linear dependence in a spatial coordinate.
These backgrounds could be interpreted as tilted spatially
homogeneous models where the normal describing the flow of 
matter, in this case the dilaton, is not orthogonal to the
hypersurface of spatial homogeneity \cite{tilt}. In all other cases
the dilaton and the antisymmetric tensor field strength $H=dB$,
are spatially homogeneous. Thus these are orthogonal 
spatially homogeneous models.
In \cite{batkeh} similar relations between  different
Bianchi types were found. However, there string cosmologies
with a dilaton and antisymmetric tensor field strength
were used as seed backgrounds. Consequently, different types
of relations between Bianchi backgrounds emerged.

\begin{table}
\begin{center}
\begin{tabular}{|c|c|c|c|}
\hline
Bianchi type & $e^{-U}$ & $Z$ & Relationship\\
\hline\hline 
II & $a_1a_2$ & $\frac{a_2}{a_1}+iz$ & II $\stackrel{(\ref{Bu_x})}{\rightarrow}$ I\\
\hline
IV & $a_2a_3e^{2x}$ & $\frac{a_2a_3}{a_2^2+a_3^2\left(f+x\right)^2}-
i\frac{a_3^2\left(f+x\right)}{a_2^2+a_3^2\left(f+x\right)^2}$ &
IV$\stackrel{(\ref{Bu_y})}{\rightarrow}$ VI$_{-1}$\\
\hline
V & $a_2a_3e^{2x}$ & $\frac{a_3}{a_2}$ &V$\stackrel{(\ref{Bu_x})}{\rightarrow}$ VI$_{-1}$\\
& & &V $\stackrel{(\ref{Bu_y})}{\rightarrow}$ VI$_{-1}$\\
\hline
VI$_{-1}$ & $a_1a_2$ &$\frac{a_2}{a_1}e^{2x}$ & VI$_{-1}$ $\stackrel{(\ref{Bu_x})}{\rightarrow}$
V\\
\hline
LRS VIII & $a_1a_3\cosh y$ &
$\frac{a_1a_3\cosh y}{a_1^2\cosh^2 y +a_3^2\sinh^2 y}-i
\frac{a_3^2\sinh y}{a_1^2\cosh^2y+a_3^2\sinh^2 y}$ &
LRS VIII $\stackrel{(\ref{Bu_y})}{\rightarrow}$ KS (open)\\
\hline
LRS IX & $a_1a_3\cos y$ & 
$\frac{a_1a_3\cos y}{a_1^2\cos^2 y+a_3^2\sin^2 y}+i\frac{a_3^2 \sin y}
{a_1^2\cos^2 y +a_3^2 \sin^2 y}$ &
LRS IX  $\stackrel{(\ref{Bu_y})}{\rightarrow}$ KS (closed)\\
\hline
\end{tabular}
\caption{Bianchi backgrounds and their duals. 
The second and third column give the functions $U$ and $Z$
of the Bianchi model in the first column.
The last column denotes
to which Bianchi model a given Bianchi model is related to
using either the T-duality transformation (\ref{Bu_x}) or (\ref{Bu_y}).
KS (open/closed) denotes the Kantowski-Sachs model with open or closed spatial
sections. The last entry was already noted in \cite{flvm}. Further details
are given in the text. }
\end{center}
\label{tab1}
\end{table}

The structure of the dual backgrounds (\ref{Bu_x}) and (\ref{Bu_y}) 
shows that the dual of a plane
wave is again a plane wave. Choosing a null geodesic in the $(t,z)$-plane
where $u=t-z$, $v=t+z$, with $z$ a longitudinal coordinate and $t$ a 
timelike variable, the radial Penrose limit is found 
by the limiting procedure (\ref{penrl}) 
for $u\rightarrow u$, $v\rightarrow \Omega^2v$,
$x\rightarrow \Omega x$, $y\rightarrow \Omega y$.
Effectively this reduces all functions, i.e. $M(u,v)$, $Z(u,v)$ and
$U(u,v)$ to functions of $u$ only, which is equivalent to 
considering the limes $v\rightarrow 0$ \cite{alex}.
Hence obtaining first the radial Penrose limit and 
then applying abelian T-duality yields the same as dualizing first 
and then obtaining the Penrose limit of the dual space-time.
Therefore the Penrose limits of various Bianchi 
cosmologies are related by duality.

\section{Examples}

The Kasner metric describes a homogeneous but anisotropic universe.
Adapted to the G$_2$ symmetry the Kasner metric can be
written as (see for example \cite{verd})
\begin{eqnarray}
ds^2=t^{(p^2-1)/2}(dt^2-dz^2)-t^{1+p}dx^2-t^{1-p}dy^2,
\label{kas}
\end{eqnarray}
where $p$ is a constant.
Close to the initial singularity the metric (\ref{g2}) is well
approximated by a Kasner metric with space-dependent 
Kasner exponents. In this case $p$ becomes a function of $z$ 
(cf. e.g. \cite{yurts}).
Introducing null coordinates $\tilde{u}=t-z$, $v=t+z$,
taking the radial Penrose limit 
and finding an affine parameter $u$
results in the following
wave profiles,
\begin{eqnarray}
h_{mm}=\kappa_m u^{-2}
\end{eqnarray}
where $m=1, 2$ and $\kappa_m$ is constant,
$\kappa_1=-\kappa_2=-p(1-p^2)/(p^2+1)^2$ for the seed metric
(\ref{kas}), $\kappa_1=-(p+1)(p^2+p+2)/(p^2+1)^2$,
$\kappa_2=\kappa_2^{(seed)}$ for the dual space-time
(\ref{Bu_x}) and 
$\kappa_1=\kappa_1^{(seed)}$,
$\kappa_2=(p-1)(p^2-p+2)/(p^2+1)^2$ for the dual
space-time (\ref{Bu_y}).
Hence in general, the wave profiles show a 
$u^{-2}$ dependence. This 
was also found in the radial Penrose limit of the flat Friedmann-Robertson-Walker
space-time and the near horizon limit of the fundamental string \cite{blau}.

Models (\ref{g2}) for which $Z$ is real or the imaginary part is 
subleading compared to the real one evolve at late times
into the Doroshkevich-Zeldovich-Novikov (DZN) universe \cite{char}.
This is an anisotropic spatially homogeneous background with an effective null fluid  
due to gravitational waves.
The DZN line element is given by
\begin{eqnarray}
ds^2=e^{2t}(dt^2-dx^2)-t^{q+1}dy^2-t^{1-q}dz^2
\label{dzn}
\end{eqnarray}
where $q$ is a constant. 
Choosing null coordinates $\tilde{u}=t-x$, $v=t+x$ 
taking the radial Penrose limit, finding the affine parameter $u$ 
the wave profiles $h_{mm}$ are
obtained as follows,
\begin{eqnarray}
h_{mm}=\alpha_m u^{-2}\left(\ln u\right)^{-2}
\left[\kappa_m+\ln u\right],
\end{eqnarray}
where $m=1, 2$ and $\alpha_m$ and $\kappa_m$ is constant,
$\alpha_1=(q+1)/2$, $\kappa_1=(1-q)/2$,
$\alpha_2=(1-q)/2$, $\kappa_2=(1+q)/2$ for the seed metric (\ref{dzn}).
$\alpha_1=-(q+1)/2$, $\kappa_1=(q+3)/2$
and $\alpha_2=\alpha_2^{(seed)}$, $\kappa_2=\kappa_2^{(seed)}$ 
for the dual space-time
(\ref{Bu_x}). $\alpha_1=\alpha_1^{(seed)}$,
$\kappa_1=\kappa_1^{(seed)}$ 
and $\alpha_2=(q-1)/2$, $\kappa_2=(3-q)/2$
for the dual
space-time (\ref{Bu_y}).

There are a few known nonsingular solutions with G$_2$ symmetry 
(cf. \cite{ns, nd, mars}).
Since in view of the T-duality transformations nondiagonal solutions are of 
particular interest the nondiagonal solution given in \cite{nd, mars}
will be investigated. 
In the vacuum case, the line element can be written as, using the coordinates of
\cite{mars} 
\begin{eqnarray}
ds^2=e^{a^2r^2}\cosh(2at)(dt^2-dr^2)-r^2\cosh(2at)d\varphi^2
-\frac{1}{\cosh(2at)}(dz+ar^2d\varphi)^2,
\label{nons}
\end{eqnarray}
where $a$ is a constant.
It is interesting to note that whereas
the T-duality transformation with respect to $\partial_x$ leads to
another non-singular background, the T-duality transformation
with respect to $\partial_y$ leads to a singular background.
Furthermore, it will be shown that the solution (\ref{nons}) can be
generated from an already known diagonal solution.
The T-duality transformation (\ref{Bu_x})  
leads to 
\begin{eqnarray}
ds^2&=&\cosh(2at)\left[e^{a^2r^2}\left(dt^2-dr^2\right)
-dz^2-r^2d\varphi^2\right]\nonumber\\
\phi&=&\ln\cosh(2at) \;\;\;\;\;\;\;\;\;\;\;\;\;\;\;\;\;\;\;\;\;\;
B_{z\varphi} = ar^2.
\label{nons_x}
\end{eqnarray}
This is the solution in the string frame. 
Applying the conformal transformation 
$g_{\mu\nu}\rightarrow e^{-\phi}g_{\mu\nu}$ yields the action
to be the Einstein-Hilbert action.
Therefore in the Einstein frame the metric is given by
\begin{eqnarray}
^{(E)}ds^2=e^{a^2r^2}\left(dt^2-dr^2\right)
-dz^2-r^2d\varphi^2.
\label{barn1}
\end{eqnarray}
This metric belongs to a class of solution given in 
\cite{barn}. The Weyl scalars show that this solution
is regular \cite{mars}. The matter contents 
of the solution \cite{barn} is a stiff perfect fluid.
In low energy string theory a pure dilaton solution 
reduces in the Einstein frame to  general relativity 
coupled to a massless scalar field which effectively behaves
as a stiff perfect fluid. Therefore the stiff perfect solutions of
\cite{barn} can be interpreted as pure dilaton solutions in the
Einstein frame. This can be connected to the solution (\ref{nons_x})
that contains a dilaton and the antisymmetric tensor field
by using an additional symmetry of low energy string theory.

In the low energy action the antisymmetric tensor field
does not appear itself but only its field strength $H=dB$.
The equation of motion for $H$ can be solved
in four dimensions in terms of the gradient of a scalar field,
the axion, $b$ . With this $H$ is given by
$H^{\mu\nu\lambda}=e^{2\phi}\epsilon^{\rho\mu\nu\lambda}b_{,\rho}$
\cite{str-co}.
Introducing 
a complex scalar field $\lambda=b+ie^{-\phi}$ it can be 
shown \cite{sen} that the equations of motion 
in the Einstein frame are invariant under an 
$SL(2,\mathbf{R})$ transformation,
$$\lambda\rightarrow\frac{\alpha\lambda+\beta}
{\gamma\lambda+\delta} \;\;\;\;\;\;\;\;\;\;\;\;
\alpha\delta-\beta\gamma=1, \;\;\;\;\;\;\alpha,\beta,\gamma,\delta\in R.
$$ The Einstein frame metric stays invariant under this transformation.
It follows, in particular, that a solution containing
a dilaton and an axion can be obtained out of 
a pure dilaton solution. 
A dilaton linear in time is a viable source for the 
stiff perfect fluid space-time (\ref{barn1}). 
Therefore performing an $SL(2,\mathbf{R})$ transformation on $\lambda
=ie^{-2at}$ leads 
to $e^{\phi}=\cosh(2at)$ and $b=\tanh(2at)$ which gives
$B_{z\varphi}=ar^2$ which is exactly (\ref{nons_x}).
Therefore the nondiagonal metric (\ref{nons}) can be reduced to the 
diagonal perfect fluid solution of \cite{barn}.
In this case, the symmetries of string cosmology have been used
as solution generating techniques in standard relativity.
This is possible due to the fact that a massless scalar field 
behaves as a stiff perfect fluid.

Whereas the application of the abelian T-duality transformation (\ref{Bu_x})
leads to a nonsingular background, this is not the 
case for the application of (\ref{Bu_y}).

The Penrose limit and the resulting plane waves are found 
by introducing the null coordinates $u=t-r$, $v=t+r$.
The final expressions are given in terms of the
non-affine paramter $u$.
For the solution (\ref{nons}) the amplitudes are given by
\begin{eqnarray}
h_{11}&=&a^2
\frac{e^{-\frac{a^2}{2}u^2}}
{\cosh^4(au)}
\left[-3\cosh^2(au)-au\sinh(au)\cosh(au)+6\right]
\nonumber\\
h_{12}&=&a^2
\frac{e^{-\frac{a^2}{2}u^2}}
{\cosh^4(au)}
\left[au\cosh(au)+6\sinh(au)\right]
\label{nons-wave}
\end{eqnarray}
and $h_{22}=-h_{11}$.
The amplitudes are regular everywhere.
The radial Penrose limit of (\ref{nons_x}) results in a plane wave with
profile
\begin{eqnarray}
h_{11}&=&a^2\frac{e^{-\frac{a^2}{2}u^2}}
{\cosh^4(au)}
\left[au\sinh(au)\cosh(au)
+\cosh^2(au)-3\right]\nonumber\\
h_{22}&=&a^2\frac{e^{-\frac{a^2}{2}u^2}}
{\cosh^4(au)}
\left[au\cosh(au)+3\sinh(au)\right]\sinh(au).
\label{nons-wavex}
\end{eqnarray}
The wave profile is regular everywhere.
The expressions for the amplitudes of the dual
wave obtained from applying abelian T-duality 
with respect to $\partial_y$ (\ref{Bu_y}) are rather lengthy and
are given in the Appendix.
The different wave profiles are shown in figure 1.
\begin{figure}
\centerline{\epsfxsize=1.65in\epsfbox{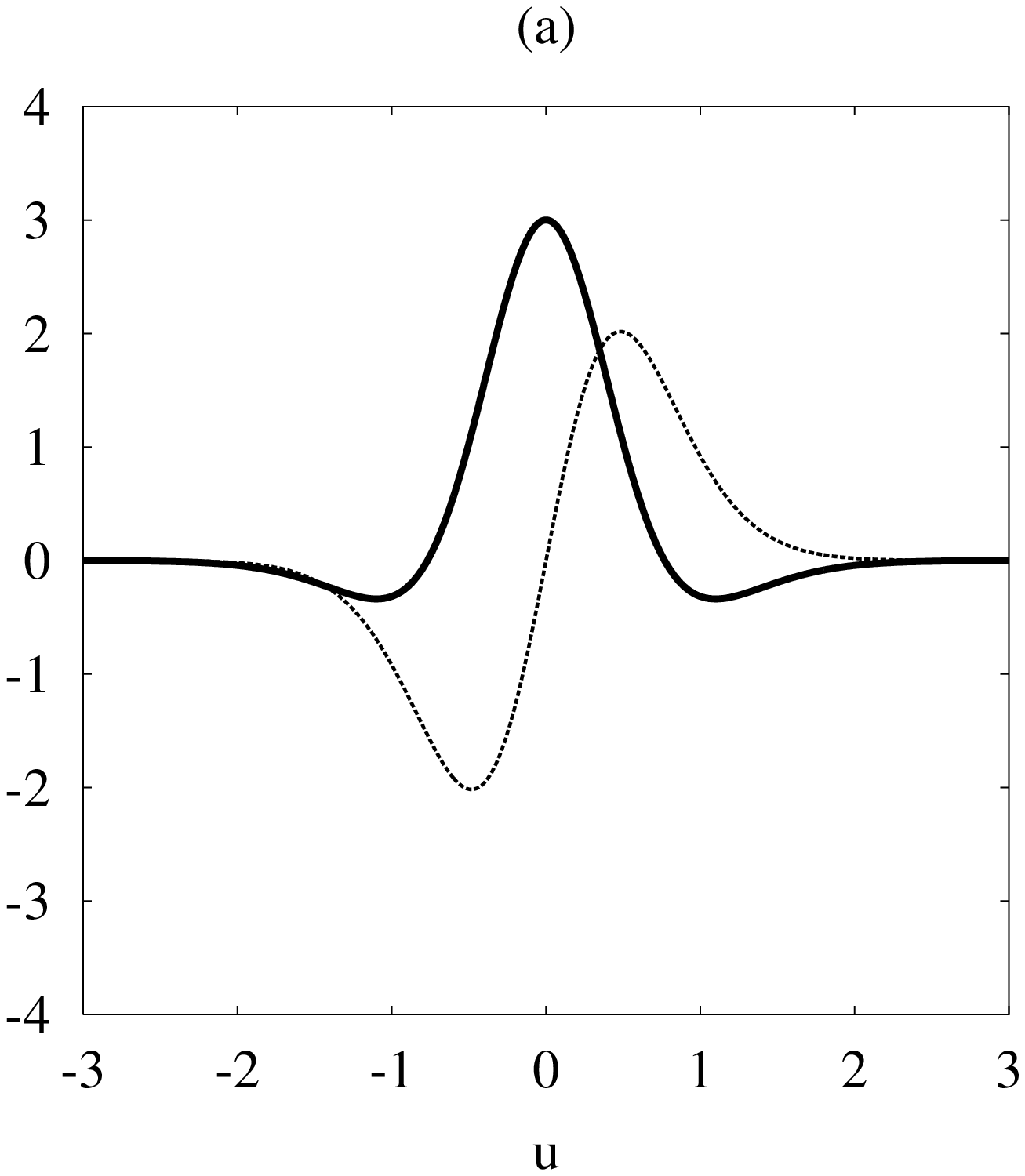} \hspace{0.05cm}
              \epsfxsize=1.65in\epsfbox{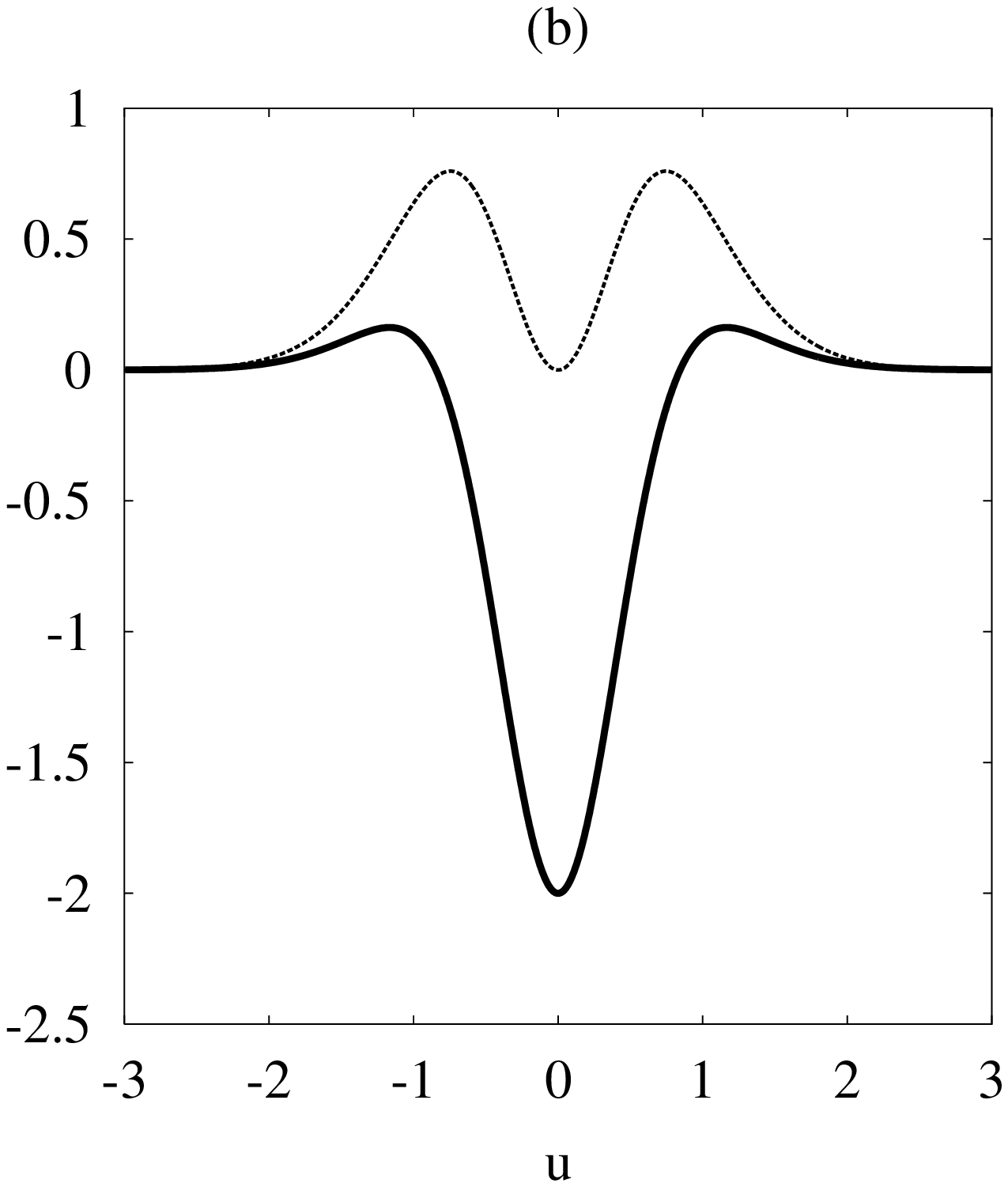}
\hspace{0.05cm}
              \epsfxsize=1.65in\epsfbox{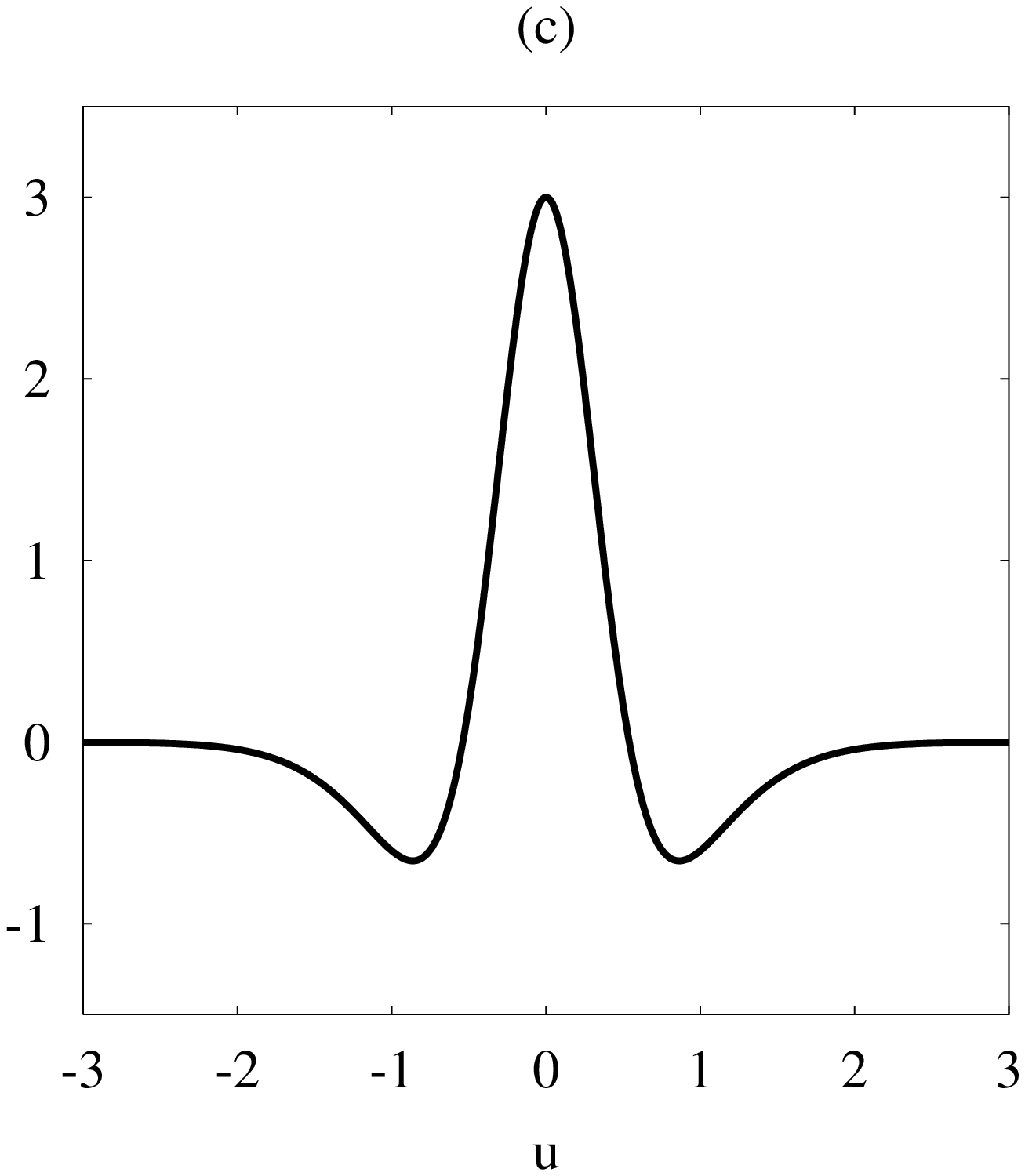}
\hspace{0.05cm}
              \epsfxsize=1.65in\epsfbox{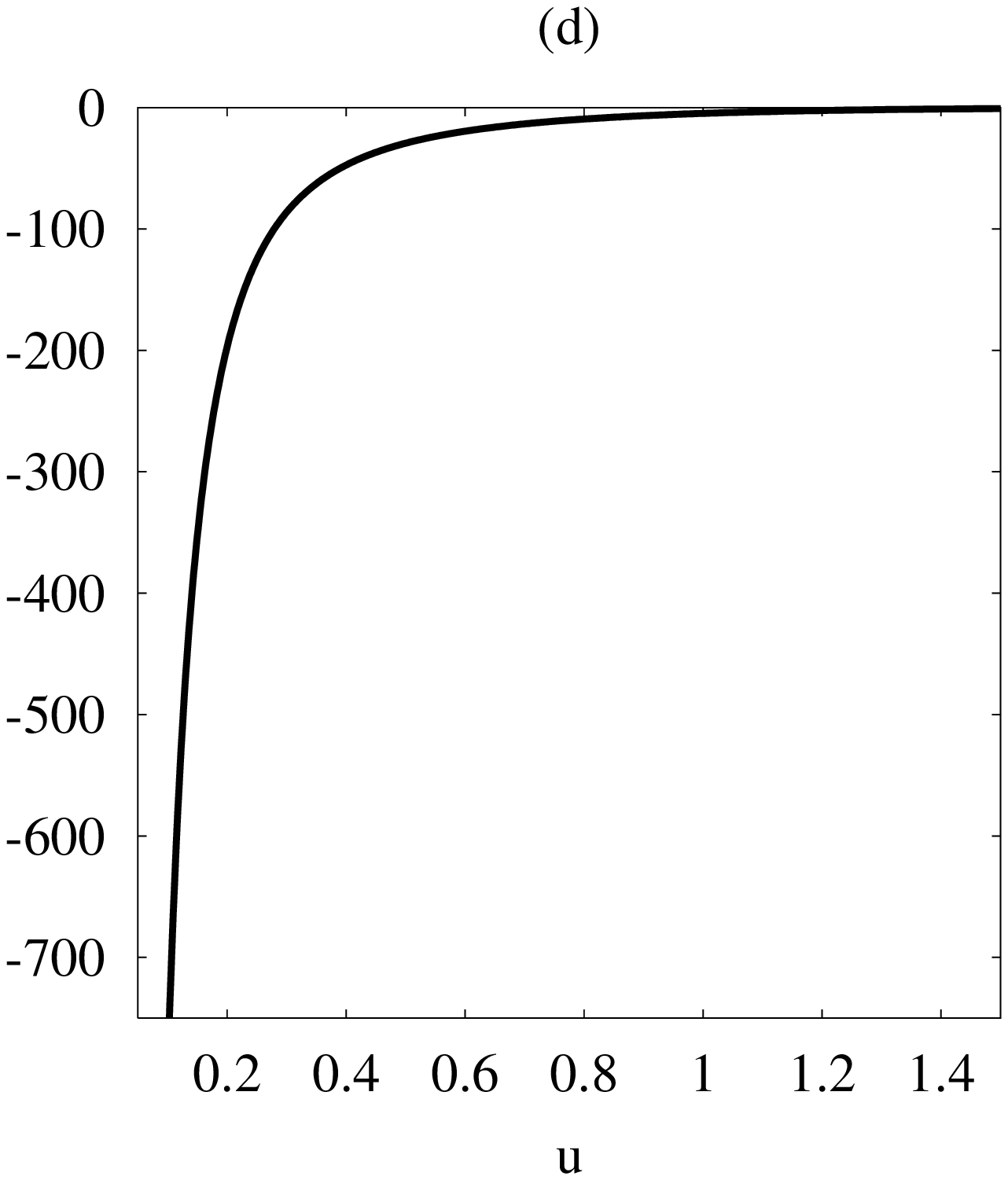}
}
\caption{(a) gives the wave profiles $h_{11}$ (black) and $h_{12}$ 
(grey) of the
plane wave in the radial Penrose limit of the
nonsingular nondiagonal metric (\ref{nons-wave}).
(b) is the wave profile of the plane wave found by duality with respect to $\partial_x$
(\ref{nons-wavex}) ($h_{11}$ (black), $h_{22}$ (grey)). (c) shows  $h_{11}$ 
and (d) $h_{22}$ of the 
wave profile of the plane wave found by duality with respect to $\partial_y$
as given in the Appendix (\ref{dualy-wave}). $a=1$ for all figures.
}
\label{fig1}
\end{figure}
It is interesting to note that only $h_{22}$ becomes
singular for $u\rightarrow 0$ for the wave obtained from the
T-duality transformation (\ref{Bu_y}). Close to the 
singularity at $u=0$ $h_{22}$ behaves as
$h_{22}\sim u^{-2}$. 
This causes  a 
strong curvature singularity to develop.
In the approach to the singularity the string coupling $g^2=e^{\phi}$ 
diverges as $u^{-2}$.
Therefore the expansion to lowest order in the
string coupling is no longer valid.

\section{Non-abelian T-duality}
The T-dual with respect to a non-abelian group of isometries is found
by gauging the two-dimensional $\sigma$-model action, integrating 
over the introduced gauge fields and gauge fixing the obtained action \cite{oq}.
Before gauge-fixing this leads to a dual action of the form \cite{oq}
in the notation of \cite{grv}
\begin{eqnarray}
S'=S+\frac{1}{4\pi}\int d^2z\left(A^{\gamma}
\bar{u}_{\gamma}
+\bar{A}^{\delta}u_{\delta}
+A^{\gamma}m_{\gamma\delta}\bar{A}^{\delta}
\right),
\label{spri}
\end{eqnarray}
where
\begin{eqnarray}
u_{\delta}&=&-\partial\tilde{X}_{\delta}+\partial X^M
\left(G_{MN}+B_{MN}\right)\xi^{N}_{\delta}\nonumber\\
\bar{u}_{\gamma}&=&\bar{\partial}\tilde{X}_{\gamma}
+\xi^{M}_{\gamma}\left(G_{MN}+B_{MN}\right)\bar{\partial}X^N\nonumber\\
m_{\gamma\delta}&=&C^{\lambda}_{\gamma\delta}\tilde{X}_{\lambda}
+\xi^M_{\gamma}\left(G_{MN}+B_{MN}\right)\xi^{N}_{\delta},
\end{eqnarray}
where greek indices are group indices and latin indices are target
space-time indices. $G_{MN}$ is the metric on the target space-time
and $B_{MN}$ the antisymmetric tensor field. $S$ is the original
$\sigma$-model action (\ref{sig}). $\tilde{X}_{\lambda}$ are
Lagrange multipliers introduced to keep the gauge connection
flat. Depending on the gauge fixing they can become coordinates
in the T-dual background.

To investigate whether taking the Penrose limit and dualizing
the background space-time are commuting procedures we need to
find the non-abelian T-dual of a plane wave background.
It is convenient to consider the metric of a plane wave in the
following form,
\begin{eqnarray}
ds^2=2dudv-e^{-U}\left(e^V\cosh Wdx^2-2\sinh W dxdy+e^{-V}\cosh Wdy^2\right),
\end{eqnarray}
where $U$, $V$ and $W$ are functions of $u$ only.
This metric admits  5 Killing vectors \cite{grif},
\begin{eqnarray}
\xi_1=\partial_x \hspace*{2.0cm}
\xi_2=\partial_y \hspace*{2cm}
\xi_3=\partial_v \hspace*{2cm}\nonumber\\
\xi_4=x\partial_v+P_{-}(u)\partial_x+N(u)\partial_y \hspace*{1.5cm}
\xi_5=y\partial_v+P_{+}(u)\partial_y+N(u)\partial_x,
\label{pw}
\end{eqnarray}
where 
$P_{\pm}(u)=\int e^{U\pm V}\cosh W du$,
$N(u)=\int e^U\sinh W du$. All commutators vanish except for
$\left[\xi_1, \xi_4\right]=\xi_3$ and $\left[\xi_2, \xi_5\right]=\xi_3$.
With $\left[\xi_{\alpha}, \xi_{\beta}\right]=C^{\mu}_{\alpha\beta}\xi_{\mu}$, the 
only non-vanishing group structure constants are given by
$C^3_{14}=1=C^{3}_{25}$. There are two semi-simple subgroups,
${\cal G}_1=\left\{\xi_1, \xi_3, \xi_4\right\}$ and 
${\cal G}_2=\left\{\xi_2, \xi_3, \xi_5\right\}$.
In the following non-abelian T-duality with respect to the subgroup 
${\cal G}_1$ will be considered.
Furthermore, it will be assumed that the dilaton 
and the antisymmetric tensor field vanish, i.e.
$\phi\equiv 0$, $B_{MN}\equiv 0$.

The first step to find the non-abelian dual with respect to the subgroup
${\cal G}_1$,
following the procedure of \cite{oq},
is to calculate the matrix $m$. It is found that $m$ is given by
\begin{eqnarray}
m=\left(
\begin{array}{ccc}
G_{xx}&0&\tilde{X}_3+G_{xx}P_{-}+G_{xy}N\\
0&0&0\\
-\tilde{X}_3+P_{-}G_{xx}+NG_{xy} &  0 & G_{xx}P_{-}^2+2NP_{-}G_{xy}+
G_{yy}N^2
\end{array}
\right).
\end{eqnarray}
The null Killing vector $\xi_3$ leads to a singular part in the T-dual action.
This yields a singular space-time that is singular everywhere if 
one tried to integrate over the gauge fields. 
Something similar happens in the case of abelian T-duality if the 
isometry has a fixed point \cite{pp}. 
In the case of the Euclidean two dimensional black hole 
the horizon, on which the time-like Killing vector becomes null, 
is interchanged with a curvature singularity in the 
T-dual background \cite{bh}.
It can also be seen in a straightforward manner in the example of
the T-dual of a two dimensional plane \cite{rv}, 
\begin{eqnarray}
ds^2=dr^2+r^2d\theta^2.
\end{eqnarray}
The T-dual with respect to the isometry $T=\partial_{\theta}$
is given by 
\begin{eqnarray}
ds^2=dr^2+r^{-2}d\theta^2.
\end{eqnarray}
The dilaton is given by $\phi=-\ln r^2$. 
The background becomes singular at $r=0$ which is exactly the point at which
$T^2=0$.
In the case of the plane wave space-time the Killing vector
$\xi_3$ is null everywhere. Even though the other two Killing vectors of
${\cal G}_1$ are not null
the T-dual space-time is singular everywhere.
Furthermore the T-dual dilaton is given by \cite{oq,grv}
\begin{eqnarray}
\phi'=\phi-\log {\rm det}\; m
\end{eqnarray}
which is singular everywhere in the T-dual background since ${\rm det}\; m=0$.

Both non-abelian subgroups, ${\cal G}_1$ and ${\cal G}_2$,
of the group 
of motions of a simple plane 
wave space-time contain one null Killing vector. Therefore 
using the procedure of \cite{oq} to find the
non-abelian T-dual of a pure plane wave results in a singular
T-dual background. 
Nevertheless, 
since 
the effective metric is built out of $G_{MN}$ and $B_{MN}$,
taking a non-vanishing
antisymmetric tensor field $B_{MN}$ 
into account might lead to 
a T-dual background that is not singular everywhere.
However, a constant $B-$ field is not enough since its Lie derivatives
in the direction of the Killing vectors of the isometry group
in general do not vanish. In that case, further terms have to be
taken into account in the T-dual action (\ref{spri}) \cite{nond1}.

Another possibility to find non-abelian T-duals of a plane wave 
that are not singular everywhere arises if the plane wave 
space-time admits additional (non-null) isometries. For example, the
WZW model of \cite{nw} admits an additional non-semisimple 
group. Non-abelian T-duals with respect to these
group have been found in \cite{tse} and \cite{moh}.
In both cases it was found that  non-abelian 
T-duality transforms the original plane wave space-time
into a background  that  is not a plane wave.

Other examples, of plane wave space-times with additional 
space-like isometries are the solutions of \cite{ht} which admit
Bianchi type IV. However, since the Bianchi IV is a non-semi-simple
group it is not possible to use the procedure of \cite{oq}.
In that case one would have to apply Poisson-Lie T-duality 
to find an equivalent solution \cite{pois}.

Therefore, in general, taking the Penrose limit and then taking the
non-abelian T-dual or first taking the non-abelian T-dual and then 
the Penrose limit results in completely different backgrounds.
This will be illustrated with the example of a vacuum Bianchi II 
cosmology.
Its metric is given by
\begin{eqnarray}
ds^2=-dt^2+a_1^2(dx-zdy)^2+a_2^2dy^2+a_3^2dz^2,
\label{bII}
\end{eqnarray}
where $a_i=a_i(t)$ \cite{taub}. 
The only non-vanishing group structure 
constant is $C^1_{23}=1$ \cite{exact}.

In \cite{grv} the non-abelian T-duals of spatially homogeneous backgrounds
have been found.
The transformed metric, antisymmetric tensor field
and shifted dilaton are given by 
\begin{eqnarray}
\tilde{G}=(\gamma-\beta-\kappa)^{-1}\gamma(\gamma+\beta+\kappa)^{-1}
\hspace*{3cm}\nonumber \\
\tilde{B}=-(\gamma-\beta-\kappa)^{-1}(\beta+\kappa)(\gamma+\beta+\kappa)^{-1}
\hspace*{1cm}
\tilde{\phi}=\phi-{\rm log\; det}(\kappa+\gamma+\beta)
\label{nosp}
\end{eqnarray}
where $\kappa$ is an antisymmetric matrix defined by
$\kappa_{\alpha\beta}\equiv C_{\alpha\beta}^{\gamma}\tilde{X}_{\gamma}$.
$\tilde{X}^{\lambda}$ are coordinates in the dual space-time.
$\gamma_{\mu\nu}(t)$ is the metric in the invariant basis on hypersurfaces
of constant time, $ds^2=-dt^2+\gamma_{\mu\nu}(t)\omega^{\mu}\omega^{\nu}$,
and $\beta_{\mu\nu}(t)$ describes the antisymmetric tensor field
in the synchronous frame $B=\beta_{\mu\nu}(t)\omega^{\mu}\wedge\omega^{\nu}$.
Furthermore $d\omega^a=\frac{1}{2}C^{\alpha}_{\mu\nu}\omega^{\mu}\wedge\omega^{\nu}$.
For Bianchi type A models $\gamma_{\mu\nu}(t)$ is diagonal,
namely, $\gamma_{\mu\nu}(t)={\rm diag}\left(a_1^2(t), a_2^2(t), a_3^2(t)\right)$.

Applying the non-abelian T-duality transformation (\ref{nosp})
to the Bianchi II vacuum background (\ref{bII})
yields to
\begin{eqnarray}
ds^2&=&-a_1^{-2}\left(d\eta^2-dx^2\right)+
\left[
 \left(a_2a_3\right)^2+x^2
\right]^{-1}
\left(
 a_3^2dy^2
 +a_2^2dz^2
\right)\nonumber\\
\phi&=&-\ln\left[\left(a_1a_2a_3\right)^2+a_1^2x^2\right],
\hspace*{2cm}
B_{yz}=-\frac{x}{\left(a_2a_3\right)^2+x^2}.
\label{b2nad}
\end{eqnarray}
This metric is no longer of Bianchi type II.
It admits two abelian Killing vectors $\partial_y$,
$\partial_z$.
Introducing null coordinates $u=\eta-x$,
$v=\eta+x$, the radial Penrose limit is found to be
\begin{eqnarray}
d\hat{s}^2&=&-\frac{1}{a_1^2}dudv+
\left[
 \left(a_2a_3\right)^2+\frac{u^2}{4}
\right]^{-1}
\left(
a_3^2dy^2
 +a_2^2dz^2
\right)
\nonumber\\
\hat{\phi}&=&-\ln\left[\left(a_1a_2a_3\right)^2+a_1^2
\frac{u^2}{4}\right],
\hspace*{2cm}
\hat{B}_{yz}=\frac{u}{2\left(a_2a_3\right)^2+
\frac{u^2}{2}}
\label{nad_ii}
\end{eqnarray}
where $a_i=a_i(u)$.

Next we will consider the radial Penrose limit of the Bianchi II cosmology
(\ref{bII}) written in the form (\ref{g2}).
The metric (\ref{bII}) admits the following three Killing vectors \cite{exact}
\begin{eqnarray}
\xi_1=\partial_x\hspace*{2cm}
\xi_2=\partial_y\hspace*{2cm}
\xi_3=\partial_z+y\partial_x.
\end{eqnarray}
Introducing null coordinates $u=\eta-z$, $v=\eta+z$, rescaling 
according to (\ref{resc}) and finding the limit $\lim_{\Omega\rightarrow 0}
\Omega^{\Delta_{\xi}}\xi_{\alpha}(\Omega)$, where $\Delta_{\xi}\in {\mathbf R}$
\cite{blau},
it turns out that, whereas $\xi_1$ and $\xi_2$ stay unchanged, $\xi_3$
becomes the null Killing vector $\partial_v$.
In addition, there are the two Killing vectors $\xi_4$ and $\xi_5$.
Hence, there are no additional isometries to the pure
plane wave isometries (\ref{pw}). Therefore, the non-abelian T-dual 
can only be found with respect to one of the subgroups, 
${\cal G}_1$ or ${\cal G}_2$, respectively. As was shown above, this 
leads to a dual background that is singular everywhere.
However, the Penrose limit of the non-abelian T-dual 
of the vacuum Bianchi II cosmology (\ref{nad_ii})
only becomes singular locally.
Thus, taking the Penrose limit and finding the non-abelian T-dual
are not commutative procedures.

Finally, some comments on non-abelian T-duality as a solution
generating technique will be made.
In the last section it was shown that abelian T-duality can be
used to connect solutions to general relativity of varying
degree of generality. Basically, starting with one solution 
a more general solution was found. In general relativity the 
approach to the initial singularity is still an open question
(for a recent account, see \cite{ug}) which is 
partly due to the fact that there are no known
general solutions. The majority of known solutions admits 
some kind symmetries. However, due to the nature of 
non-abelian T-duality most of the symmetries of the original
space-time will be broken in the T-dual background.
Therefore, one might use these transformations 
to generate very general solutions which admit, if at all,
only few isometries. This will be discussed with the example
of Bianchi VIII and IX as seed metrics.
These are the most general spatially homogeneous metrics. 
Furthermore, their group structure is semi-simple. 
The group structure
constants are given by, $C^{1}_{23}=\pm 1$, $C^2_{31}=1$, $C^3_{12}=1$,
where the upper sign corresponds to Bianchi IX and the lower one
to Bianchi VIII. 
The non-abelian T-duality transformation (\ref{nosp}) yields to
\begin{eqnarray}
\tilde{G}&=&\left(a_1^2a_2^2a_3^2+a_1^2x^2+
a_2^2y^2+a_3^2z^2\right)^{-1}\left(
	\begin{array}{ccc}
        a_2^2a_3^2+x^2 & \pm xy & \pm xz\\
	\pm xy & a_1^2a_3^2+y^2 & yz\\
	\pm xz  & yz & a_1^2a_2^2+z^2
	\end{array}
        \right)\\
\tilde{B}&=&\left(a_1^2a_2^2a_3^2+a_1^2x^2+
a_2^2y^2+a_3^2z^2\right)^{-1}\left(
	\begin{array}{ccc}
	0 & a_3^2z-2\epsilon xy & -a_2^2y-2\epsilon xz\\
	-a_3^2z+2\epsilon xy & 0 & \pm a_1^2x\\
	a_2^2y+2\epsilon xz & \mp a_1^2 x& 0
	\end{array}
        \right) \\
\tilde{\phi}&=&-\log\left(a_1^2a_2^2a_3^2+a_1^2x^2+a_2^2y^2+a_3^2z^2
\right),
\end{eqnarray}
where $\epsilon=1$ for Bianchi IX and $\epsilon=0$ for Bianchi VIII.
This background  
could be interpreted as an inhomogeneous generalization of
a Bianchi I background. For small values of $x$, $y$ and $z$ the spatial
part imposes a small perturbation on a Bianchi I background.
The vacuum Bianchi IX metric with three different scale factors
is the Mixmaster model which shows chaotic behaviour. However,
the evolution of the scalefactors can be approximately described by 
a succession of Kasner epochs, each of them determined by
a set of Kasner exponents $(\alpha_1, \alpha_2, \alpha_3)$ (cf. e.g.\cite{mix}). 
The Kasner metric is given by $ds^2=-dt^2+t^{2\alpha_1}dx^2+
t^{2\alpha_2}dy^2+t^{2\alpha_3}dz^2$, and, in vacuum, the exponents
satisfy $\sum_i\alpha_i=1=\sum_i\alpha_i^2$.
Using such a solution in the expressions for the 
scale factors of the seed vacuum Bianchi IX metric one finds
that the initial singularity persists in the 
T-dual background. Furthermore the metric is approximately diagonal,
with the new scale factors being $1/a_i$. Hence there
will be also Mixmaster oscillations in the dual background, though 
due to the presence of the scalar field and the antisymmetric 
tensor field
these will cease after 
a finite number of oscillations \cite{dh}. Furthermore close to the 
singularity the T-dual universe enters into a strongly coupled 
regime, since the string coupling $g^2=e^{\phi'}$ diverges
for $t\rightarrow 0$. 

The T-dual background is very inhomogeneous though it is also rather special,
since the spatial dependence is completely fixed and does not 
allow for arbitrary constants, as it is the case for the scale factors $a_i(t)$.

\section{Conclusions}
In the Penrose limit any space-time in the vicinity of a null
geodesic can be 
approximated by a plane wave.
Since plane waves are exact classical string vacua this might 
help to connect cosmological solutions to an underlying 
string vacuum. There are only very few known exact solutions 
that have a cosmological interpretation and these  are very far away
from describing our observable universe.
Since plane waves are classical string vacua it makes sense to 
find a first quantized theory of a string propagating in these
backgrounds. 
This has been studied in particular for singular backgrounds
with wave profiles following a power law in the null
coordinate \cite{str-g}. Here it was found that this type of wave profile
occurs for the radial Penrose limit of a Kasner universe,
whose scale factors are following  a power law in cosmic time.
For space-times with more general functional behaviour
different types of evolution were found. In particular the
wave profiles of the plane wave obtained in the radial
Penrose limit of a nonsingular cosmological solution
were determined. Here it might be interesting to study the
first quantization of a string propagating in this background.

Low energy string theory admits a number of symmetries. 
These have been used to find more
exact solutions, their corresponding Penrose limits
and wave profiles.
In addition relationships between different 
spatially homogeneous backgrounds have been found.
This is interesting from the point of view that
abelian T-duality and taking the radial Penrose limit
are commuting procedures. 
Furthermore using abelian T-duality and the 
$SL(2,\mathbf{R})$ invariance of low energy string theory
it was found that the nonsingular nondiagonal
solution \cite{nd, mars} can be reduced to a diagonal
static solution \cite{barn}. This shows once 
more that the symmetries of low
energy string theory can be used to learn more 
about solutions in general relativity.

The non-abelian T-dual of a vacuum 
plane wave space-time has been investigated in 
detail. It was found that if there are no additional
isometries then dualizing with 
respect to one of the semi-simple subgroups of
isometries of the plane wave leads to a T-dual
background that is singular everywhere. The reason for that 
is the presence of a null  Killing vector in each subgroup.
This is similar to what happens in abelian T-duality.
If there are additional isometries one might 
find T-dual backgrounds that are not singular everywhere.
However, due to the nature of non-abelian T-duality,
the T-dual of these kind of  plane waves
will not be a plane wave. There are some known
examples. It might also be interesting to
discuss these issues in the context of Poisson-Lie T-duality. 
Therefore, since the Penrose limit leads to a plane wave space-time,
taking the Penrose limit and applying a non-abelian
T-duality transformation are, in general, not commutative
procedures. On the contrary, taking the Penrose
limit and applying abelian T-duality are commutative.

Finally, the role of abelian and non-abelian T-duality as
solution generating techniques has been discussed.
In particular,
non-abelian T-duality was used to find 
more general inhomogeneous solutions which can be interpreted
as inhomogeneous generalizations of a Bianchi I cosmology.

\section*{Acknowledgements}
It is a pleasure to thank E. \'Alvarez, J.L.F. Barb\'on,
A. Feinstein and especially 
M.A. V\'azquez-Mozo
for enlightening discussions.
I am grateful to J.L.F. Barb\'on for discussions
on T-duality.

\section*{Appendix: The profile of the dual wave obtained from (\ref{nons}) with 
(\ref{Bu_y})}
The dual wave profile resulting from applying
(\ref{Bu_y}) to the background (\ref{nons})
is given by
\begin{eqnarray}
h_{11}&=&a^2e^{-\frac{a^2}{2}u^2} 
\left[\frac{a^5u^5\sinh(au)\cosh(au)
-a^4u^4\left(\cosh^2(au)+3\right)
+16a^2u^2\cosh^2(au)\left(2\cosh^2(au)-3\right)}
{\cosh^4(au)\left(4\cosh^2(au)+a^2u^2\right)^2}
\right.
\nonumber\\
&+&\left.
\frac{-16au\sinh(au)\cosh^3(au)\left(\cosh^2(au)+6\right)
+48\cosh^4(au)\left(2-\cosh^2(au)\right)}
{\cosh^4(au)\left(4\cosh^2(au)+a^2u^2\right)^2}
\right]
\nonumber\\
h_{22}&=&e^{-\frac{a^2}{2}u^2} 
\left[
\frac{a^7u^7\sinh(au)\cosh(au)-3a^6u^6\left(\cosh^2(au)+1\right)
+16a^4u^4\cosh^2(au)\left(\cosh^2(au)-4\right)}
{u^2\cosh^4(au)\left(4\cosh^2(au)+a^2u^2\right)^2}
\right.
\nonumber\\
&-&\left.
\frac{16a^3u^3\sinh(au)\cosh^3(au)\left(8+\cosh^2(au)\right)
80a^2u^2\cosh^6(au)}
{u^2\cosh^4(au)\left(4\cosh^2(au)+a^2u^2\right)^2}
\right.
\nonumber\\
&-&\left.
\frac{
128au\sinh(au)\cosh^5(au)+128\cosh^6(au)}
{u^2\cosh^4(au)\left(4\cosh^2(au)+a^2u^2\right)^2}
\right].
\label{dualy-wave}
\end{eqnarray}

\end{document}